\documentclass[acmtog]{acmart}
\usepackage[
  cachedir=minted-cache
]{minted}
\usepackage{seqsplit}
\usepackage{subcaption}
\usepackage{multirow}
\usepackage{array}
\usepackage{graphicx}
\usepackage{tikz}
\usetikzlibrary{arrows.meta, positioning, calc}
\AtBeginDocument{%
  }

\newcommand{\codefont}{\fontsize{7.0pt}{8.2pt}\selectfont}

\copyrightyear{2026}
\acmYear{2026}
\setcopyright{cc}
\setcctype{by-nc-nd}
\acmConference[SA Conference Papers '26]{SIGGRAPH Asia 2026 Conference Papers}{December 01--04, 2026}{Kuala Lumpur, Malaysia}
\acmBooktitle{SIGGRAPH Asia 2026 Conference Papers (SA Conference Papers '26), December 01--04, 2026, Kuala Lumpur, Malaysia}
\acmDOI{10.1145/3829340.3842170}
\acmISBN{979-8-4007-2842-6/2026/12}

\acmSubmissionID{1143}

\begin{document}

\title{InstantMimic: A High Performance System for Learning Physics-based Skills in Seconds}

\author{Ikjun Choi}
\email{ikjun@imo.snu.ac.kr}
\orcid{0009-0004-5274-8739}
\affiliation{%
  \institution{Seoul National University}
  \city{Seoul}
  \country{South Korea}
}

\author{Geonho Leem}
\email{geonholeem@imo.snu.ac.kr}
\orcid{0009-0001-5236-4389}
\affiliation{%
  \institution{Seoul National University}
  \city{Seoul}
  \country{South Korea}
}

\author{Jungdam Won}
\authornote{corresponding author}
\orcid{0000-0001-5510-6425}
\email{jungdam@imo.snu.ac.kr}
\affiliation{%
  \institution{Seoul National University}
  \city{Seoul}
  \country{South Korea}
}

\begin{teaserfigure}
    \centering
    \includegraphics[width=\linewidth]{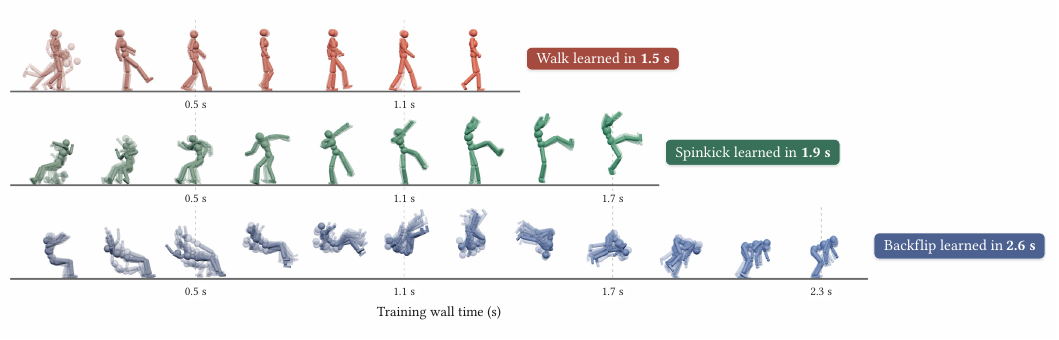}
    \caption{A learning-progress visualization for walk, spinkick, and backflip motions.}
    \Description{Three rows show training progress for walk, spinkick, and backflip. Overlaid humanoid poses follow each learned motion as training wall time advances; the policies complete the three skills in 1.5, 1.9, and 2.6 seconds, respectively.}
    \label{fig:teaser}
\end{teaserfigure}

\begin{abstract}
Physics-based character control is a long-standing challenge in computer graphics and robotics, requiring policies that satisfy complex dynamics while producing realistic motion. Recent Deep RL approaches, particularly imitation learning methods such as DeepMimic, have had broad impact beyond animation, influencing robotics by enabling agile and expressive behaviors. While these approaches achieve impressive results, they remain computationally inefficient to train in practice. Despite GPU-accelerated simulation, we find that end-to-end pipelines often underutilize hardware due to overheads outside the physics solver, caused by fragmented GPU kernels and CPU memory access in the critical path.
We present InstantMimic, a system that addresses these inefficiencies by making the entire training loop GPU-native. Built on a GPU-native physics backend, our unified pipeline integrates simulation, environment computation, policy inference, and policy updates within a single execution flow. As a result, InstantMimic reduces training time for diverse physics-based skills to a few seconds and makes LLM-agent-driven hyperparameter search practical.
\end{abstract}



\maketitle

\section{Introduction}

Physics-based character control has long been a central problem in computer graphics and robotics, enabling simulated agents to produce realistic and physically consistent motions. Unlike kinematic approaches, physics-based methods must satisfy complex dynamic constraints, making control significantly more challenging while offering greater realism and generalization.

To address this challenge, a line of work has leveraged deep reinforcement learning (Deep RL) to learn control policies directly from motion data. In particular, imitation learning frameworks such as DeepMimic~\cite{peng2018deepmimic} have demonstrated that neural policies can reproduce a wide range of dynamic skills, including locomotion, acrobatics, and transitions, by tracking reference motions. These approaches have significantly advanced the state of the art in physics-based animation, enabling robust and versatile controllers without requiring manual controller design. Beyond computer graphics, their impact has extended to robotics, where Deep RL-based imitation policies have become a foundation for learning agile and expressive humanoid behaviors~\cite{ze2025twist2, liao2025beyondmimic}.

Despite these advances, efficient training of such systems remains elusive. Recent GPU-accelerated physics simulators have enabled substantial speedups for reinforcement learning, yet simply executing the physics engine on the GPU does not guarantee efficient end-to-end learning. In practice, we observe that the GPU is often underutilized during training. Through profiling of a representative motion imitation pipeline, we find that the primary bottlenecks do not lie in the physics solver itself, but in the surrounding training pipeline. 
In particular, two sources of overhead consistently dominate runtime: \emph{GPU kernel fragmentation}, caused by excessive fine-grained execution without sufficient batching or fusion, and \emph{CPU memory access in the critical path}, caused by Python-level abstractions.
Importantly, these inefficiencies are not specific to a particular task or framework, 
but commonly emerge in Deep RL training loops developed without end-to-end profiling and systems-level optimization.
As a result, even state-of-the-art implementations fail to fully utilize available hardware, leaving substantial performance untapped.

In this work, we revisit the design of the Deep RL training loop for physics-based character control from a systems perspective. We focus on eliminating the systemic inefficiencies that hinder existing approaches. Our key idea is to restructure the entire training loop to be fully GPU-native, minimizing kernel launch overhead through batching and fusion, and avoiding implicit synchronization. We design a unified execution model in which simulation, observation construction, reward evaluation, and policy updates are tightly integrated within a training loop, ensuring continuous device utilization throughout training.

As a result, our approach enables a dramatic reduction in training time. Our system learns policies that reproduce diverse physics-based skills within only a few seconds of training, while maintaining high motion fidelity and robustness. Beyond raw speed, this shift fundamentally changes how such systems are used.
When training completes in seconds, manual hyperparameter tuning becomes the bottleneck: a human cannot iterate fast enough to exploit the speedup. An LLM agent offers two things manual tuning lacks: iteration at machine speed and unattended operation across many cycles, exploring the hyperparameter space and reliably discovering high-performing configurations. 
This transforms physics-based skill learning from a slow, trial-and-error process into a rapid and largely automated workflow.

We summarize our contributions as follows:
\begin{itemize}
    \item \textbf{A system-level analysis of inefficiencies in the Deep RL training loop} for physics-based character control, identifying data movement, CPU-GPU synchronization, and kernel launch overhead as key bottlenecks.

    \item \textbf{Seconds-level learning of physics-based skills}, reducing training time from minutes or hours to only a few seconds.
    
    \item \textbf{Enabling automated experimentation}, including LLM agent-based hyperparameter optimization that iterates at machine speed with minimal human intervention.
    
    \item \textbf{An open-source training framework} to support reproducibility and future research. Code and data are available at \url{https://github.com/Scripter36/InstantMimic}.
\end{itemize}

\begin{figure*}[h!]
  \centering
  
\begin{subfigure}[t]{1.0\linewidth}
    \centering
    \includegraphics[width=\linewidth,trim={0 0.5cm 0 0cm}]{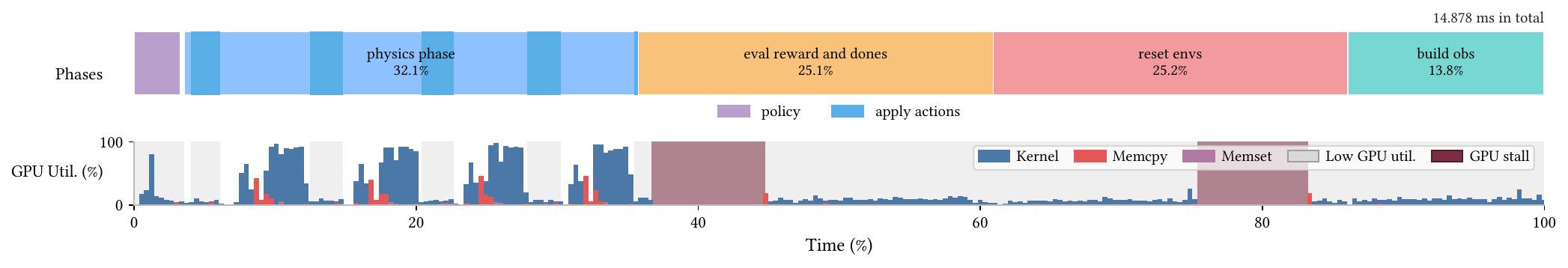}
    \caption{Isaac Lab-based training loop (baseline).}
    \label{fig:nsys_step_isaaclab}
\end{subfigure}
    \hfill
\begin{subfigure}[t]{1.0\linewidth}
    \centering
  \includegraphics[width=\linewidth,trim={0 0.5cm 0 0cm}]{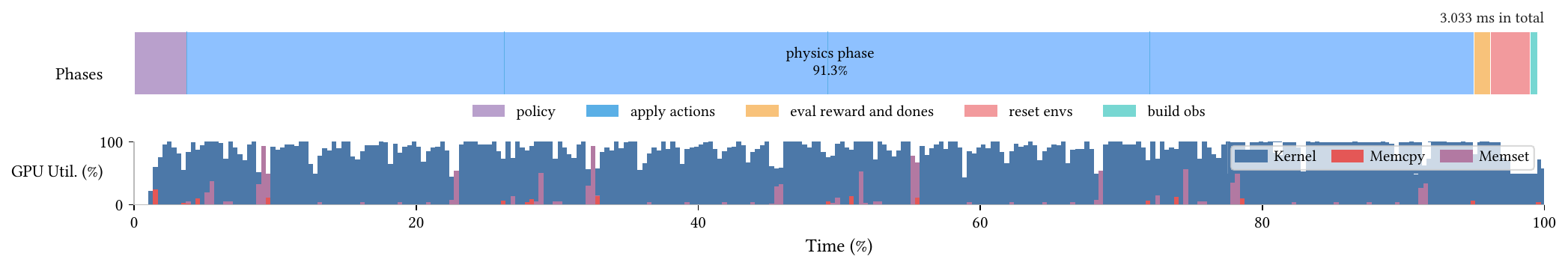}
    \caption{GPU-native training loop (ours).}
    \label{fig:nsys_step_ours}
\end{subfigure}

  \caption{%
    GPU profiling timeline of a single rollout step in two PPO training loops.
  }
  \Description{Two vertically stacked timelines decompose one rollout step into phases and plot GPU utilization over normalized time. The Isaac Lab loop takes 14.878 milliseconds and contains long low-utilization and stalled regions during reward, reset, and observation work. The GPU-native loop takes 3.033 milliseconds, spends 91.3 percent of the step in the physics phase, and keeps GPU utilization high through most of the step.}
  \label{fig:nsys_step}
\end{figure*}

\section{Related Work}

\subsection{GPU-accelerated Physics Simulation}

Physics-based character control has long relied on CPU-based rigid body simulators such as ODE~\cite{ode} and MuJoCo~\cite{todorov2012mujoco}, which supported trajectory optimization and early deep RL in motor skill learning~\cite{hodgins1995animating, yin2007simbicon, sok2007simulating, coros2010generalized, lee2010data, peng2018deepmimic, bergamin2019drecon, won2020scalable}. As RL matured, on-policy algorithms such as PPO requiring billions of environment interactions made simulation throughput the binding bottleneck. Isaac Gym~\cite{makoviychuk2021isaac} and Brax~\cite{freeman2021brax} addressed this by performing rigid body dynamics in parallel on the GPU, reducing humanoid controller training from days to minutes~\cite{rudin2022learning}. This throughput in turn enabled skill-learning frameworks trained on large collections of motion data~\cite{peng2021amp, peng2022ase, tessler2023calm, tessler2024maskedmimic}. Related efforts in differentiable and deformable simulation, including Taichi-based systems~\cite{hu2019taichi, hu2019chainqueen}, explored similar GPU-centric execution strategies for particle and continuum dynamics. More recently, MuJoCo Warp~\cite{mujoco_warp_2025}, built on NVIDIA Warp~\cite{warp2022}, brings this level of parallelism to the MuJoCo stack. We build our framework on MuJoCo Warp.

\subsection{Optimization Principles in GPU Computing}
The massive parallelism of modern GPUs has driven large speedups in deep RL training. However, realizing these speedups in practice requires addressing three classes of overhead that are well-established in the parallel computing literature~\cite{kirk2016programming}.

\paragraph{Kernel launch overhead.}
A GPU kernel is a function-like program that is launched by the CPU and executed in parallel by many GPU threads. Every kernel launch carries a fixed CPU-side dispatch cost on the order of microseconds. For short-running kernels, such as small matrix multiplications, this cost can approach or exceed the kernel's own execution time, leaving the GPU idle between launches. \textit{Kernel fusion}, facilitated by JIT compilation frameworks such as Warp~\cite{warp2022}, merges multiple operations into a single kernel, reducing both the number of launches and intermediate memory traffic. \textit{CUDA Graphs}~\cite{nvidia_cuda_graphs} instead capture a sequence of launches once and replay it with a single API call, eliminating per-launch CPU overhead without changing the kernels themselves. The two are orthogonal, one a compile-time transformation and the other a runtime batching mechanism.

\paragraph{Data movement.}
On modern systems, data movement, not arithmetic, dominates latency~\cite{horowitz2014computings}. The principle is to access data in place rather than copy it, both across the CPU--GPU boundary and within device memory. End-to-end GPU simulators such as Isaac Gym~\cite{makoviychuk2021isaac} and Brax~\cite{freeman2021brax} apply this zero-copy discipline by keeping rollout state resident on the device. The same discipline applies to \emph{every} stage of the training loop such as policy inference, reward computation, and logging.

\paragraph{Synchronization.}
CPU--GPU data transfers are also the most common trigger of implicit synchronization: reading a GPU scalar on the CPU forces the CPU to wait for all preceding GPU work to complete, stalling the GPU pipeline even when the transfer itself is small. CUDA streams and events enable asynchronous CPU--GPU execution, but this benefit is lost when a blocking access is issued, often through a high-level tensor API~\cite{makoviychuk2021isaac}.

Madrona~\cite{shacklett2023extensible} applies these principles to batched environment simulation in a programmable GPU-native framework, primarily on game-like environments, while leaving policy inference and optimization to an external learning framework. In contrast, InstantMimic targets high-DoF humanoid motion imitation with full 3D articulated-body physics and integrates the entire training loop, including simulation, policy inference, and PPO optimization into a GPU-native framework.

\subsection{Hyperparameter Optimization for Deep RL}

The performance of a deep RL policy is sensitive to the hyperparameters used during training, and automated hyperparameter optimization (HPO) has therefore been a research topic since the early stages of machine learning~\cite{bergstra2012random}. Since hyperparameters are not differentiable with respect to the training loss, most HPO methods treat training as a black box. Bayesian optimization (BO), which fits a probabilistic surrogate over past evaluations~\cite{jones1998efficient, shahriari2015taking}, is a representative example and serves as the backbone of production HPO services such as Google Vizier~\cite{golovin2017google} and Amazon SageMaker~\cite{perrone2021amazon}. 
For deep RL specifically, where the optimal hyperparameters are non-stationary over the course of training, Population Based Training (PBT)~\cite{jaderberg2017population} has been proposed to adapt a schedule of hyperparameters over the course of learning.

These methods, however, share a structural limitation: the number of evaluations required to reach a target performance grows rapidly with the dimensionality of the search space~\cite{wang2016towards}, forcing researchers in practice to restrict optimization to a small, hand-picked subset of hyperparameters. Designing such a search space well is crucial for achieving high performance~\cite{bergstra2012random}. However, search space design is still largely driven by human expertise.

Recent work has demonstrated that LLMs can act as a proxy for human expertise in configuring training. Some methods prompt an LLM to propose hyperparameters directly or use an LLM as a surrogate inside BO~\cite{zhang2023using, liu2024llambo}, reducing the search cost over wide configuration spaces. More open-ended approaches go further by allowing LLM agents to directly modify training code~\cite{karpathy2026autoresearch, chan2024mle, ferreira2026can}. Because this process can be automated, LLM-based methods can explore configuration spaces more extensively than approaches that rely on human expertise.

\section{Profiling Bottlenecks in the Baseline}
\label{sec:analysis}

\begin{figure}[t]
  \centering
  \begin{minted}[breaklines,frame=lines,framesep=2mm,fontsize=\codefont]{python}
while training:
    for _ in range(horizon): # rollout steps
        action = policy(obs)

        # physics phase
        for _ in range(num_physics_substeps):
            apply_action(action)
            simulate_physics()

        # post physics phase
        reward, terminated, truncated = eval_reward_and_dones()
        reset_envs(terminated | truncated)
        storage.write(obs, action, reward, terminated, truncated)
        obs = build_obs()

    update_policy(storage)
  \end{minted}
  \caption{%
    Pseudocode of the PPO training loop.
  }
  \Description{Pseudocode shows PPO training as a rollout loop followed by a policy update. Each rollout step performs policy inference, repeated physics substeps, reward and termination evaluation, environment reset, rollout storage, and observation construction.}
  \label{fig:rl_training_loop}
\end{figure}

Profiling a DeepMimic-style training loop on Isaac Lab~\cite{mittal2025isaac} reveals that, even with a GPU-based physics engine, the GPU is stalled or underutilized for $77\%$ of the wall time per rollout step (Figure~\ref{fig:nsys_step_isaaclab}). 

To localize where these bottlenecks arise, we decompose a standard RL training loop~\cite{towers2026gymnasium} shown in Figure~\ref{fig:rl_training_loop}. Each step of the rollout loop, from policy evaluation to rollout-storage write, is a \emph{rollout step}. Within each rollout step, the \emph{physics phase} covers action application and simulator substeps, and the \emph{post-physics phase} covers the remaining work (reward, reset, observation, storage updates).

We identify two root causes for the observed bottlenecks: (i) fragmented post-physics kernels that prevent the GPU from staying busy between physics phases, and (ii) CPU memory access stalls that disrupt continuous GPU execution.

\subsection{Low GPU Utilization from GPU Kernel Fragmentation}
One might expect physics computation to dominate the \emph{rollout step} wall time. However, profiling tells a different story. The \emph{post-physics phase} accounts for $64\%$ of each rollout step, exceeding the \emph{physics phase} at \(32\%\) (the sum of orange, red, cyan blocks vs. the blue block in Figure~\ref{fig:nsys_step_isaaclab}). Even within the \emph{physics phase}, \mintinline{python}{apply_action} alone consumes roughly \(29\%\) of this phase. Consequently, the physics simulation itself accounts for only \(22.3\%\) of the full rollout step. Worse, outside physics simulation, the GPU is active for only $7.3\%$ of the total wall time on average, well below full utilization. The GPU therefore sits underutilized for the majority of each rollout step, despite every operation being GPU-resident in principle.

We attribute this underutilization to \emph{kernel fragmentation}, 
where computation is dispatched as many short GPU kernels separated by CPU launch gaps, 
rather than a few large kernels. 
In the training loop, fragmentation arises from the implementation pattern. All \emph{rollout step} work outside the physics simulation is commonly written as a sequence of PyTorch tensor operations invoked from the Python interpreter, and each operation dispatches its own GPU kernel even when it performs only a small amount of work. For such fragmented kernels, the kernel launch overhead becomes comparable to, or larger than, the GPU execution time, so the GPU frequently waits for the CPU to enqueue the next operation. 
The resulting timeline (gray block in Figure~\ref{fig:nsys_step_isaaclab}) shows persistently low GPU utilization during these stages, consistent with fragmented short-kernel execution and CPU-side launch overhead.

\subsection{CPU Memory Access in Critical Path Stalls the GPU}

Another source of inefficiency is visible in the burgundy regions of Figure~\ref{fig:nsys_step_isaaclab}, where the GPU remains stalled during CPU–GPU memory exchanges. Whenever data must be exchanged between CPU and GPU memory, the GPU cannot proceed until the transfer completes, and the CPU cannot proceed until the GPU has produced or consumed the data. Each such exchange therefore stalls the GPU on the critical path of the \emph{rollout step}. Python abstractions hide these costs behind a tensor-like interface, leaving them invisible at the call site; profiling reveals them at a glance. We illustrate with two representative patterns from the Isaac Lab baseline.

The state accessor API of Isaac Lab \texttt{body\_com\_pose\_b} (Figure~\ref{fig:isaaclab_state_accessors}), invoked during \mintinline{python}{build_obs}, reads from a CPU-resident simulation buffer and copies the result to the GPU on every call, aligning with an approximately \(15.9\,\%\) GPU stall in Figure~\ref{fig:nsys_step_isaaclab}.

The \texttt{reset} method stalls the GPU through a transfer in the opposite direction, from GPU back to CPU. Termination flags are computed on the GPU, but the Python branch must read a scalar back to the CPU via \texttt{.item()} before launching reset (Figure~\ref{fig:cpu_side_branch}). The lone \texttt{.item()} call is easy to overlook in code review yet forces a full GPU sync, placing a synchronization point on the \emph{rollout step}.

\section{GPU-Native Training Loop}
\label{sec:method}

We built a GPU-native training loop that addresses the bottlenecks identified in Section~\ref{sec:analysis}. To reduce kernel fragmentation, we apply kernel fusion and CUDA Graph replay techniques across the \emph{post-physics phase} and PPO updates. To eliminate CPU memory access, we keep state access, reset handling, and training summaries on the GPU. This loop is built on MuJoCo Warp, a GPU-native MuJoCo backend that exposes simulator state as GPU arrays, replacing Isaac Lab used in the baseline.

\begin{figure}[H]
    \centering
    \begin{subfigure}[t]{1.0\linewidth}
        \centering
    \begin{minted}[breaklines,frame=lines,framesep=1.5mm,fontsize=\codefont]{python}
@property
def body_com_pose_b(self) -> torch.Tensor:
    # Below copies CPU memory into GPU
    pose = self._root_physx_view.get_coms().to(self.device)
    pose[..., 3:7] = math_utils.convert_quat(pose[..., 3:7], to="wxyz")
    return pose
        \end{minted}
        \caption{Simulation state accessors.}
        \label{fig:isaaclab_state_accessors}
    \end{subfigure}

    \medskip

    \begin{subfigure}[t]{1.0\linewidth}
        \centering
    \begin{minted}[breaklines,frame=lines,framesep=1.5mm,fontsize=\codefont]{python}
terminated = compute_termination()
if terminated.any().item():
    reset(terminated)
        \end{minted}
        \caption{Conditional reset.}
        \label{fig:cpu_side_branch}
    \end{subfigure}

    \caption{%
        Two CPU memory access patterns in Isaac Lab.
    }
    \Description{Two code examples show CPU access on the rollout critical path. The first copies center-of-mass poses from CPU to GPU, and the second reads a GPU termination flag on the CPU before resetting environments.}
    \label{fig:cpu_stalls}
\end{figure}

\subsection{Reducing GPU Kernel Fragmentation}
\label{sec:kernel_overhead}
\begin{figure}[H]
  \centering
  \begin{subfigure}[t]{\linewidth}
    \begin{minted}[breaklines,frame=lines,framesep=1.5mm,fontsize=\codefont]{python}
def tracking_reward_torch(...):
    pose_err = (body_pos - ref_body_pos).square().sum(dim=(-1, -2))
    vel_err = (body_vel - ref_body_vel).square().sum(dim=(-1, -2))
    com_err = (com_pos - ref_com_pos).square().sum(dim=-1)

    reward = (0.60 * torch.exp(-4.0 * pose_err)
            + 0.25 * torch.exp(-0.25 * vel_err)
            + 0.15 * torch.exp(-8.0 * com_err))
    done = (root_height < 0.55) | (reward < 0.15)

    reward_out.copy_(reward)
    done_out.copy_(done.to(torch.int32))
    \end{minted}
    \caption{PyTorch-based reward computation.}
  \end{subfigure}

  \vspace{0.25em}
  \begin{subfigure}[t]{\linewidth}
    \begin{minted}[breaklines,frame=lines,framesep=1.5mm,fontsize=\codefont]{python}
@wp.kernel
def tracking_reward_warp(...):
    env = wp.tid()
    pose_err = float(0.0)
    vel_err = float(0.0)

    for body in range(NUM_BODIES):
        dp = body_pos[env, body] - ref_body_pos[env, body]
        dv = body_vel[env, body] - ref_body_vel[env, body]
        pose_err += wp.dot(dp, dp)
        vel_err += wp.dot(dv, dv)

    dc = com_pos[env] - ref_com_pos[env]
    reward = (0.60 * wp.exp(-4.0 * pose_err)
            + 0.25 * wp.exp(-0.25 * vel_err)
            + 0.15 * wp.exp(-8.0 * wp.dot(dc, dc)))
    reward_out[env] = reward
    done_out[env] = int(root_height[env] < 0.55 or reward < 0.15)
    \end{minted}
    \caption{NVIDIA Warp-based reward computation.}
  \end{subfigure}

  \vspace{0.25em}
  \begin{subfigure}[t]{\linewidth}
    \centering
    \includegraphics[width=\linewidth]{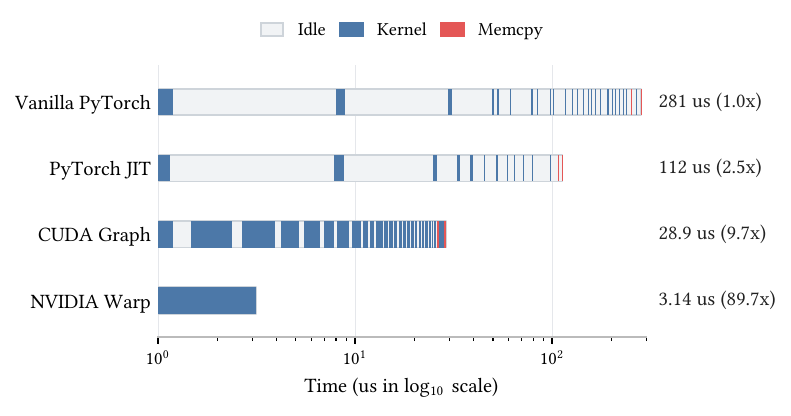}
    \caption{GPU kernel profile for vanilla PyTorch, PyTorch JIT, CUDA Graph, and Warp implementations.}
    \label{fig:reward_timing}
  \end{subfigure}

  \caption{%
    Reward computation benchmark illustrating the techniques for reducing kernel fragmentation.
  }
  \Description{The first two panels express the same reward and termination computation with PyTorch tensor operations and a single Warp kernel. The execution timelines in the third panel compare vanilla PyTorch, PyTorch JIT, CUDA Graph, and Warp; the Warp implementation has the fewest kernel launches and the shortest runtime.}
  \label{fig:reward_fusion_microbench}
\end{figure}

We illustrate the effect of reducing kernel fragmentation on reward computation with two techniques that together yield a roughly \(90\times\) speedup (Figure~\ref{fig:reward_fusion_microbench}). The first reduces launch gaps between existing kernels. The second fuses operations into larger kernels, avoiding intermediate operations and temporary tensors.

To isolate the effect of each technique, we prepare four variants of the reward computation — vanilla PyTorch (\emph{vanilla}), PyTorch JIT (\emph{jit}), CUDA Graph (\emph{cgraph}), and NVIDIA Warp (\emph{warp}). The \emph{vanilla} uses the typical PyTorch tensor operation API. The \emph{jit} partially fuses tensor operations using the PyTorch JIT API, but still leaves the reward computation split across multiple GPU kernels; this technique is used by Isaac Lab. The \emph{cgraph} uses CUDA Graphs to preserve the original kernel function bodies and launch sequence while reducing gaps between kernel launches through single-launch graph replay. The \emph{warp} variant instead uses NVIDIA Warp to JIT-compile the reward computation into a single GPU kernel, eliminating intermediate operations and approaching the fragmentation-free limit.

We break down the contribution of each technique by comparing the variants. The \emph{jit} applies partial fusion but inherits \emph{vanilla}'s per-kernel launch gaps, yielding only a \(2.5\times\) speedup (\(281 \to 112\,\mu s\)). The \emph{cgraph} keeps the same kernels but eliminates the launch gaps via single-shot graph replay, yielding a \(9.7\times\) speedup (\(29\,\mu s\)). The \emph{warp} variant, which expresses the reward as a single GPU kernel, benefits from both: no intermediate tensors (one kernel body) and no launch gaps (one launch). The result is an \(89.7\times\) speedup ($3.14\,\mu s$), nearly two orders of magnitude over \emph{vanilla}. Equivalently, the shrinking idle gaps across these variants (Figure~\ref{fig:reward_timing}) confirm the diagnosis in Section~\ref{sec:analysis}: kernel fragmentation causes low GPU utilization.

We therefore implement the \emph{post-physics phase} as explicit \emph{warp} kernels. For PPO updates we use the PyTorch-side equivalent of \emph{warp} (\texttt{torch.compile}).

\subsection{Removing CPU Memory Access from the Critical Path}
\label{sec:host_synchronization}

Eliminating CPU memory access on the \emph{rollout step} turns the stalled profile of Figure~\ref{fig:nsys_step_isaaclab} into the stall-free profile of Figure~\ref{fig:nsys_step_ours}, recovering the $16\%$ of wall time the GPU previously spent waiting on the CPU. We achieve this by relocating the two patterns identified in Section~\ref{sec:analysis}. Our GPU-native implementation reads directly from GPU-resident simulation arrays, so \emph{post-physics phase} kernels obtain simulator state without the CPU-side copy of Figure~\ref{fig:isaaclab_state_accessors}. The conditional reset of Figure~\ref{fig:cpu_side_branch} becomes unconditional; the rollout graph always executes the reset logic, and a GPU-side mask restricts the actual update to the terminated environments.

\section{LLM Agent Based Hyperparameter Optimization}
\label{sec:agentic_hpo}
Due to the high computational efficiency of our framework, evaluating controller performance under different hyperparameter configurations can be performed within seconds. This drastically reduces the cost of hyperparameter search compared to conventional approaches, where each trial may require substantial training time. Such efficiency enables a fundamentally different approach to hyperparameter optimization. Instead of relying on manual tuning or coarse search strategies, our system allows for rapid iterative experimentation, making it feasible to integrate agentic AI that autonomously explores and refines hyperparameters through repeated trials. In this setting, the agent can actively design experiments, evaluate outcomes, and adapt its strategy in a closed loop, effectively discovering high-performing configurations with minimal human intervention.

\begin{figure}[H]
  \centering
  \begin{tikzpicture}[
    node distance=0.6cm,
    stage/.style={
      draw, thick, rounded corners=3pt,
      minimum width=1.5cm, minimum height=0.7cm,
      align=center,
      font=\footnotesize,
      fill=gray!5
    },
    prompt/.style={
      draw, thick, rounded corners=3pt,
      minimum width=5.4cm, minimum height=0.85cm,
      align=center,
      font=\footnotesize,
      fill=gray!5
    },
    arr/.style={-{Latex[length=2mm]}, thick}
  ]
    \node[stage] (plan) {plan \\ {\scriptsize (LLM)}};
    \node[stage, right=of plan] (propose) {propose \\ {\scriptsize (LLM)}};
    \node[stage, right=of propose] (experiment) {experiment \\ {\scriptsize (Python executor)}};

    \coordinate (mid) at ($(plan.east)!0.5!(experiment.west)$);

    \draw[arr] (plan) -- (propose);
    \draw[arr] (propose) -- (experiment);

    \draw[arr, rounded corners=8pt]
      (experiment.south) -- ++(0,-0.4) -| (plan.south);

    \node[prompt, below=1.2cm of mid,
          label={[font=\footnotesize\itshape, align=left]left:LLM\\prompt}] (prompt) {%
      \begin{tabular}{@{}l@{\;:\;}l@{}}
        plan      & strategy, history, best parameters \\
        propose   & strategy, incumbent parameters \\
      \end{tabular}%
    };

    \end{tikzpicture}
  \caption{LLM agent hyperparameter optimization loop.}
  \Description{A cycle connects plan, propose, and experiment. The LLM updates the search strategy and proposes candidate values, while the Python executor trains and benchmarks the candidates before returning results to the next planning stage.}
  \label{fig:agentic_hpo_loop}
\end{figure}
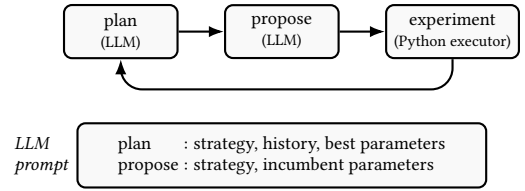

Specifically, we use an LLM agent to find hyperparameters. The agent interprets training metrics and reshapes the search space as the run progresses.
The procedure is a loop with three stages (Figure~\ref{fig:agentic_hpo_loop}). The agent maintains four artifacts as persistent state across cycles: a natural-language \emph{strategy} that records what to focus on next, a \emph{history} of prior cycles' candidates and results, the \emph{incumbent} parameters that the current search builds on, and the global \emph{best} parameters observed so far. The cycle begins with \emph{plan}, an LLM call that reads the strategy, history, and best parameters, and rewrites the strategy for the next search. \emph{propose}, a second LLM call, then reads the revised strategy and the incumbent and emits candidate values for one hyperparameter. \emph{experiment}, a deterministic stage implemented in Python, trains and benchmarks every candidate with multiple seeds, records the results, and updates the incumbent and best parameters. Further implementation details are provided in the supplemental material. The human supplies only the benchmark protocol and the stopping criterion; everything else is the agent's responsibility.

\begin{figure*}
    \centering
  \begin{subfigure}[t]{0.50562\textwidth}
    \centering
    \includegraphics[width=\linewidth]{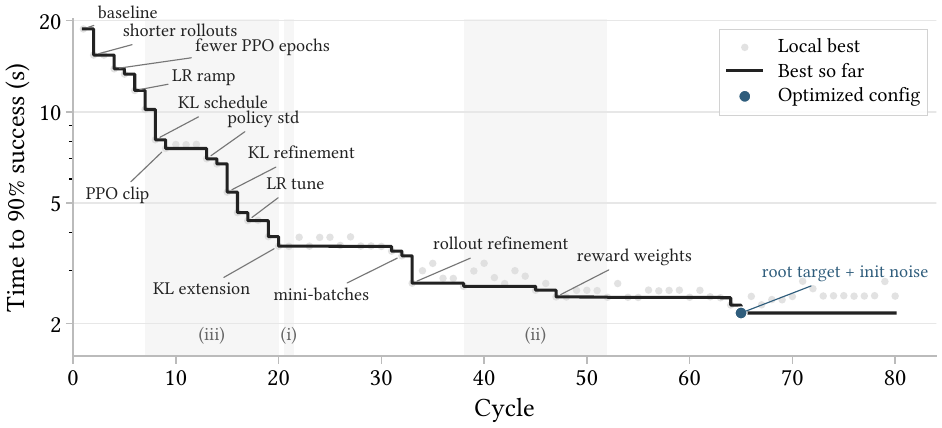}
    \caption{Optimization trajectory of the LLM agent over successive tuning cycles.}
    \label{fig:optim_traj}
  \end{subfigure}\hfill
  \begin{subfigure}[t]{0.23022\textwidth}
    \centering
    \includegraphics[width=\linewidth]{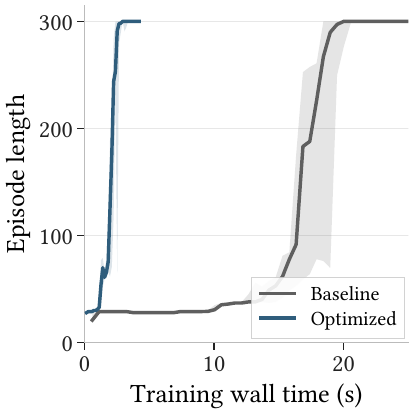}
    \caption{Learning curves for each configuration.}
  \end{subfigure}\hfill
  \begin{subfigure}[t]{0.24416\textwidth}
    \centering
    \includegraphics[width=\linewidth]{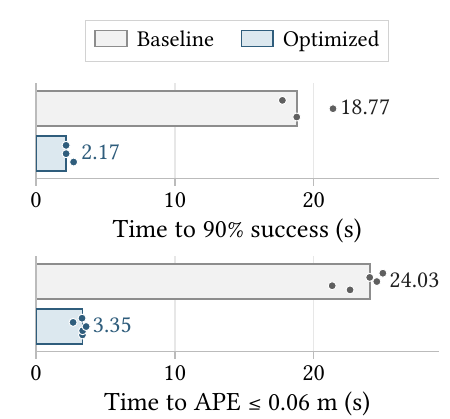}
    \caption{Training times to reach the success-rate and APE thresholds.}
  \end{subfigure}
    \caption{LLM agent hyperparameter optimization on the backflip motion tracking task. The objective is the train time required to reach a success rate over $0.9$.}
    \Description{Three panels summarize the backflip search. The first shows the time to 90 percent success decreasing across 80 cycles from 18.77 to 2.17 seconds, with major search decisions annotated. The second compares baseline and optimized learning curves. The third compares their times to 90 percent success and to an average pose error at most 0.06 meters.}
    \label{fig:backflip_autoresearch_result}
\end{figure*}

\begin{table*}[t]
  \centering
  \caption{Three representative optimization plans corresponding to the regions marked (i)--(iii) in Figure~\ref{fig:optim_traj}.}
  \label{tab:agent_decisions}
  \small
  \begin{tabular}{@{}c p{0.40\linewidth} p{0.46\linewidth}@{}}
    \toprule
    \# & Strategy & Reason \\
    \midrule
    (i)   & Select \texttt{ratio\_clip}, the PPO update-size knob most coupled to the new high-KL regime, to stabilize tracking. & Cycle~21: with \texttt{kl\_threshold} newly raised to $0.12$, success rate held at $1.0$ but average pose error and end-effector tracking error worsened. \\
    (ii)  & Pivot out of the PPO class. Reward-weight sweeps over cycles 47--52 produced a new best of $2.449\,\mathrm{s}$ at cycle 47 (\texttt{com\_reward\_weight}). & Cycles 38--46: six PPO/exploration axes (\texttt{motion\_rand\_std}, policy log-std, rollouts, action scale, future frames, prev-actions observation) had all failed to advance from a $2.656\,\mathrm{s}$ plateau. \\
    (iii) & Ratchet the bound outward each cycle, walking $0.004 \to 0.12$ ($30\times$ range). & Cycles 7--20: each winning value sat at the current upper bound of \texttt{kl\_threshold}. \\
    \bottomrule
  \end{tabular}
\end{table*}

\section{Results}
\label{sec:results}

We remove the training-loop bottlenecks (Section~\ref{sec:analysis}) with a GPU-native training loop (Section~\ref{sec:method}), and use its seconds-level training time to drive an LLM agent hyperparameter optimization (Section~\ref{sec:agentic_hpo}). This section evaluates the training loop throughput, motion tracking, and scaling to large motion datasets with downstream tasks.

\subsection{Implementation Details}
\label{sec:impl_details}

We train a humanoid character on a single NVIDIA RTX~5090 with MuJoCo Warp as the simulator. The reward is a DeepMimic-style imitation reward~\cite{peng2018deepmimic} written in multiplicative form $r = \prod_{k \in \mathcal{K}} \exp(-w_k e_k)$ over joint pose, joint velocity, center-of-mass position, and end-effector position and orientation errors. Observations follow PHC~\cite{luo2023perpetual}: root kinematics, joint state, body poses and velocities expressed in the character's heading frame using the continuous 6D orientation representation~\cite{zhou2019continuity}, future reference end-effector frames, and the previous action. Actions are normalized PD targets mapped to per-joint scales. We train with PPO, with early termination and adaptive episode-initialization sampling following~\cite{peng2018deepmimic, won2020scalable, won2019learning}.

\subsection{Training Performance for Motion Tracking}
\label{sec:training_loop_throughput}

We compare the complete training loop against two reference implementations on the backflip-tracking task: Isaac Lab with PhysX and mjlab~\cite{zakka2026mjlab} with MuJoCo Warp. With the optimizations described in Section~\ref{sec:method}, InstantMimic reaches $0.613$ million frames per second (Mfps), compared with $0.105$ Mfps for Isaac Lab ($5.85\times$) and $0.157$ Mfps for mjlab ($3.91\times$). The full-loop speedup is lower than the $89.7\times$ reward-computation speedup in Section~\ref{sec:kernel_overhead} because training also includes physics simulation, policy inference, and PPO updates.

\label{sec:hpo_results}

We next use an LLM agent (GPT-5.5-high) to find hyperparameters for a backflip motion-tracking policy, following the procedure in Section~\ref{sec:agentic_hpo}. Figure~\ref{fig:backflip_autoresearch_result} shows the resulting 80-cycle search and learning curves. Average pose error (APE) is included as a secondary tracking-quality metric. The best configuration reduces the time to success from $18.77\,\mathrm{s}$ to $2.17\,\mathrm{s}$ ($8.6\times$). Table~\ref{tab:throughput} summarizes these measurements.

\begin{table}[ht]
    \centering
    \caption{Training throughput and time to success for backflip tracking.}
    \label{tab:throughput}
    \small
    \begin{tabular}{@{}>{\raggedright\arraybackslash}p{3.6cm}>{\centering\arraybackslash}p{2.1cm}@{}}
        \toprule
        Framework & Throughput \\
        \midrule
        Isaac Lab~\cite{mittal2025isaac} & $0.105$ Mfps \\
        mjlab~\cite{zakka2026mjlab} & $0.157$ Mfps \\
        InstantMimic & $0.613$ Mfps \\
        \bottomrule
    \end{tabular}
    \begin{tabular}{@{}>{\raggedright\arraybackslash}p{3.6cm}>{\centering\arraybackslash}p{2.1cm}@{}}
        \toprule
        Training configuration & Time to success \\
        \midrule
        Initial hyperparameters & $18.77\,\mathrm{s}$  \\
        LLM-selected hyperparameters & $2.17\,\mathrm{s}$ \\
        \bottomrule
    \end{tabular}
\end{table}
\label{sec:per_motion}

\begin{figure*}
    \centering
    \includegraphics[width=\textwidth]{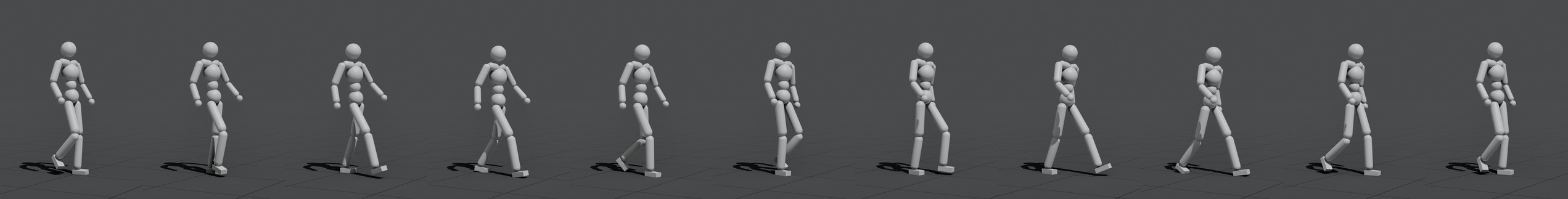} \vspace{0.05em}
    \includegraphics[width=\textwidth]{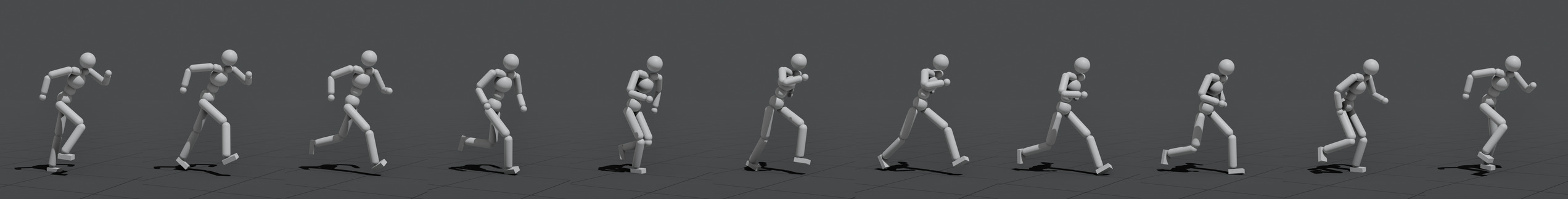} \vspace{0.05em}
    \includegraphics[width=\textwidth]{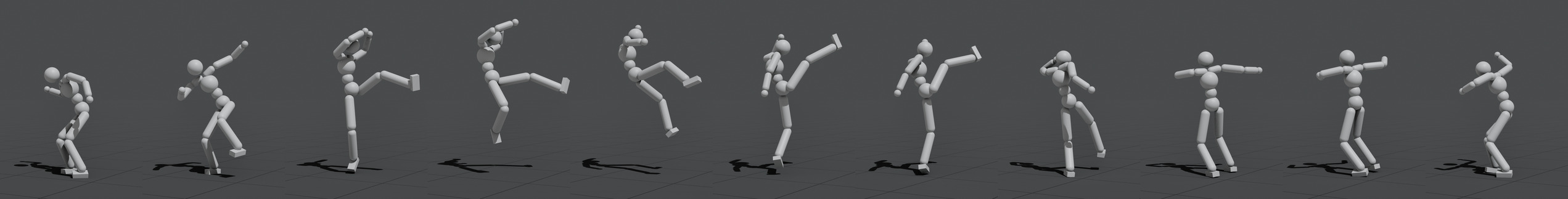} \vspace{0.05em}
    \includegraphics[width=\textwidth]{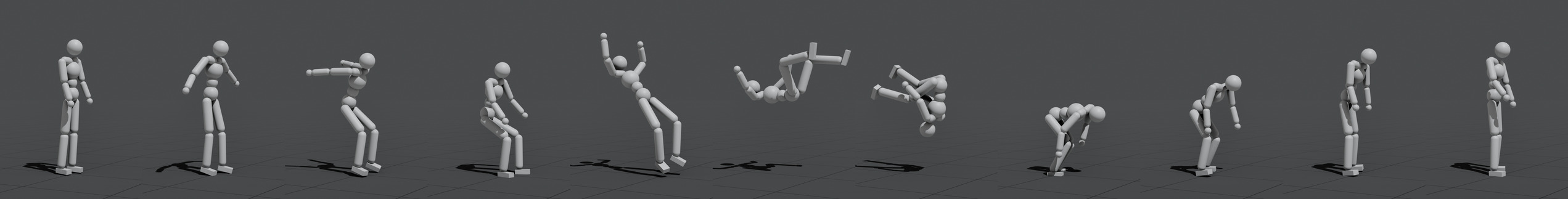} \vspace{0.05em}
    \includegraphics[width=\textwidth]{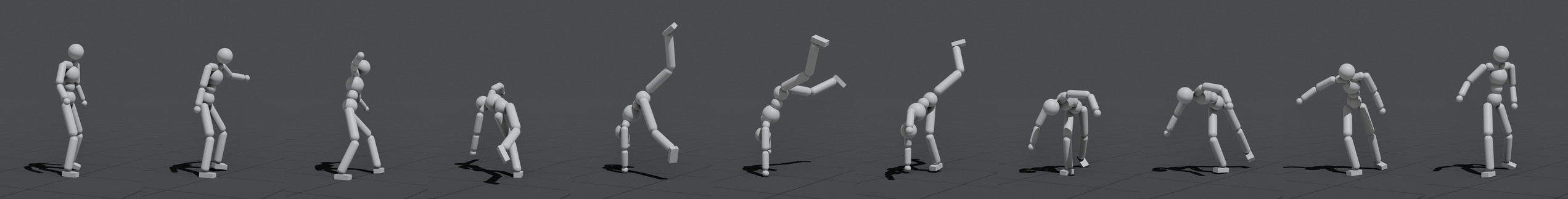}
    \caption{Qualitative motion-tracking results of five motions. From top to bottom: walk, run, spinkick, backflip, and cartwheel.}
    \Description{Five horizontal sequences show successive humanoid poses for walk, run, spinkick, backflip, and cartwheel. The acrobatic rows depict a rotating kick, an airborne backward rotation, and transitions between foot and hand contact.}
    \label{fig:motion_snapshots}
\end{figure*}

\begin{figure}[ht]
    \centering
    \includegraphics[width=\linewidth]{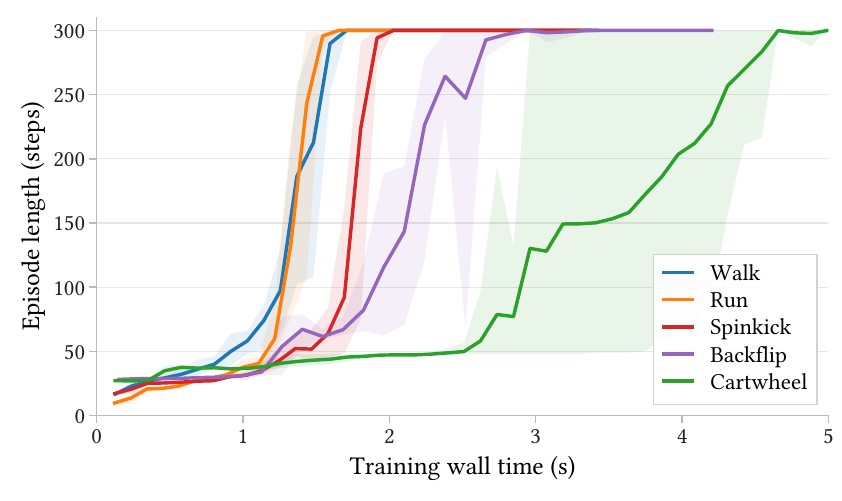}
    \caption{Learning curves for the five reference motions.}
    \Description{Episode length is plotted against training wall time for five motions, with shaded variation across seeds. Walk and run converge first, followed by spinkick and backflip; cartwheel takes the longest.}
    \label{fig:motion_learning_curves}
\end{figure}

The backflip HPO configuration described above served as the basis for the motion-specific training configurations. With these configurations, all five reference motions converge within seconds (Figure~\ref{fig:motion_learning_curves}).
The motion set covers locomotion (walk, run) and acrobatic skills
(spinkick, backflip, cartwheel); each policy reaches its tracking target
between 1.5 and 4.5 seconds of training. Representative frames from
each trained policy are shown in Figure~\ref{fig:motion_snapshots}.

\subsection{Analysis of the LLM Search}
\label{sec:hpo_analysis}

To examine the agent's decisions during the search, Table~\ref{tab:agent_decisions} catalogs three representative \emph{plan} steps corresponding to the marked regions in Figure~\ref{fig:optim_traj}, each tied to the cycle-log evidence that informed it. These exemplify three forms of search-structure modification: (i) inferring the mechanism behind a sub-metric regression, (ii) pivoting between hyperparameter classes when one is exhausted, and (iii) expanding the search bounds when winners sit at the boundary. None of these is available to a fixed-search-space optimizer such as Bayesian optimization (BO), nor to a script that sweeps a predefined grid. Such reasoning requires expert-level knowledge that fixed-search-space optimizers do not provide, applied at a per-trial pace no human can match. The LLM agent closes both gaps.

\subsection{Scaling to Large Motion Datasets}
\label{sec:amass}

\begin{figure}[ht]
    \centering
    \begin{subfigure}[t]{0.48\linewidth}
        \centering
        \includegraphics[width=\linewidth]{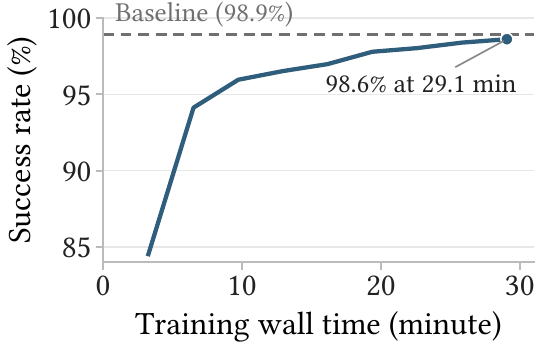}
        \caption{AMASS learning curves.}
        \label{fig:amass_large_data}
    \end{subfigure}
    \hfill
    \begin{subfigure}[t]{0.48\linewidth}
        \centering
        \includegraphics[width=\linewidth]{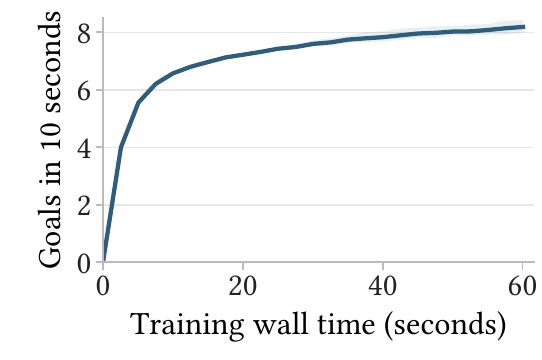}
        \caption{Goal-tracking learning curves.}
        \label{fig:amass_goal_tracking}
    \end{subfigure}

    \caption{Large-scale motion pretraining (AMASS 37.4h) and downstream goal-tracking behavior learned from the pretrained latent controller. The goal-tracking curve shows the mean and min--max range over three seeds.}
    \Description{Two learning curves show large-scale motion pretraining and a downstream task. The left curve approaches the PHC baseline success rate of 98.9 percent, reaching 98.6 percent after 29.1 minutes. The right curve rises to 8.2 goals per 10-second episode after 60 seconds, with the shaded band showing the minimum and maximum over three seeds.}
    \label{fig:amass_combined}
\end{figure}

We pretrain a VAE-based latent controller on the AMASS~\cite{mahmood2019amass} training split of PHC~\cite{luo2023perpetual} (37.4\,hours of motion), reaching a success rate comparable to the PHC baseline within 30\,minutes (Figure~\ref{fig:amass_large_data}). We then freeze the decoder and train a downstream goal-position tracking policy in its latent action space. After one minute of training, the policy reaches 8.2 goals per 10-second episode (Figure~\ref{fig:amass_goal_tracking}).

\section{Conclusion}

We revisited the design of the Deep RL training loop for physics-based character control from a systems perspective. Profiling revealed that the Isaac Lab baseline underutilized the GPU for most of each rollout step due to kernel fragmentation around physics execution and CPU-mediated memory access on the rollout critical path. By introducing optimized GPU kernels and direct GPU-memory state access, we substantially improved training throughput and reduced rollout overhead.

The resulting system enables training times on the order of seconds for standard motion-tracking tasks and reduces large-scale latent-controller pretraining on the 37.4-hour AMASS dataset to 30 minutes.

Among the five reference motions, cartwheel is the most challenging, involving inverted whole-body rotation and transitions between foot and hand contacts. It takes about $4.5$ seconds to reach the tracking target, compared with about $1.5$ seconds for walk and run. The acrobatic motions required lower component weights in the reward to accommodate larger tracking errors during learning.

These speedups also make automated experimentation significantly more practical. In particular, an LLM agent-based optimization was able to iteratively refine training configurations and reduce backflip training time from $18.77\,\mathrm{s}$ to $2.17\,\mathrm{s}$ over 80 optimization cycles.

However, the configuration found by HPO depends on how candidate performance is evaluated. Our search ranks candidates by the training time to success. At cycle 21 of the search (Figure~\ref{fig:optim_traj}), the candidate improved the primary time-to-success metric but exhibited higher tracking errors. The agent used these additional metrics to adjust its next proposal, but because they did not directly affect candidate ranking, the search still favored faster convergence over tracking quality.

Our findings suggest that, in modern GPU RL pipelines, the primary bottlenecks often lie not in simulation itself but in the surrounding training infrastructure. More broadly, we believe that aggressively reducing iteration time changes how RL systems should be developed: once experiments become cheap enough to run at scale, automated search and optimization become increasingly powerful tools for controller design. We hope our work provides a foundation for future research in this direction.

\begin{acks}
This work was supported by the National Research Foundation of Korea(NRF) grant funded by Korea government(MSIT and MOE) (No. RS-2026-25500579), and by the Institute of Information \& Communications Technology Planning \& Evaluation (IITP) grant funded by MSIT under Grant Nos. RS-2025-25442338 (AI Star Fellowship Support Program, Seoul National University), RS-2021-II211343 (Artificial Intelligence Graduate School Program, Seoul National University), and IITP-2026-RS-2020-II201460 (ITRC: Information Technology Research Center).
\end{acks}

\bibliographystyle{ACM-Reference-Format}
\bibliography{acmart}

\clearpage
\appendix
\section{Hyperparameter Optimization with an LLM Agent}
\label{sec:llm_agent_appendix}

\subsection{Agent Interface and Search Protocol}

The agent follows the \emph{plan--propose--experiment} loop described in the main paper. During the \emph{plan} and \emph{propose} stages, the Python executor constructs a stage-specific user prompt and invokes the LLM through Codex or OpenCode. We author only these prompts, which contain all required context and instruct the model not to access files or invoke tools. The LLM returns a revised strategy in Markdown or a proposal in JSON; the executor handles all subsequent state updates and experiments.

The two prompts draw on the following inputs:
\begin{itemize}
  \item \textbf{Task definition.}
  \texttt{\seqsplit{task\_description}} identifies the target skill.
  \texttt{\seqsplit{objective}} defines the ranking criterion, success threshold,
  aggregation across seeds, and timing convention.

  \item \textbf{Search space.}
  \texttt{\seqsplit{searchable\_parameters}} lists the configuration
  parameters eligible for modification and their current values.
  \texttt{\seqsplit{task\_config\_excerpt}} provides the resolved settings
  used in the \emph{propose} stage.

  \item \textbf{Search state.}
  \texttt{\seqsplit{current\_state}} records the incumbent configuration, the best
  observed result, and the most recent outcome.
  \texttt{\seqsplit{recent\_progress}} preserves recent planning decisions, while
  \texttt{\seqsplit{strategy\_excerpt}} carries the revised strategy from \emph{plan} into \emph{propose}.

  \item \textbf{Experimental evidence.}
  \texttt{\seqsplit{recent\_results}} contains outcomes from up to the 15 most recent
  cycles, and \texttt{\seqsplit{seed\_diagnostics}} shows how those outcomes vary
  across seeds. \texttt{\seqsplit{earlier\_summary}} condenses the earlier history.

  \item \textbf{Validation feedback.}
  When a proposal is rejected, \texttt{\seqsplit{validation\_feedback}} records the
  error for the next \emph{propose} call.
\end{itemize}

\definecolor{promptgray}{gray}{0.95}

\begin{figure}[ht]
  \centering
  \begin{subfigure}[t]{\linewidth}
    \centering
    \begin{minted}[breaklines,frame=lines,framesep=2mm,baselinestretch=1.05,fontsize=\scriptsize,bgcolor=promptgray,bgcolorpadding=2mm]{text}
You are planning one cycle of a bounded hyperparameter search.

Task: {task_description}
Objective: {objective}
Search space: {searchable_parameters}
Current search state: {current_state}
Recent experiments: {recent_results}
Summary of earlier experiments: {earlier_summary}
Recent planning decisions: {recent_progress}
Per-seed diagnostics: {seed_diagnostics}

Compare at least three promising parameter axes, then choose
one axis for the next experiment. Return Markdown with exactly
the following headings:

# Current Strategy
## Best Baseline
## What Seems Promising
## What To Avoid
## Next Experiment
## Wildcards
  \end{minted}
    \caption{\emph{plan} prompt.}
    \label{fig:agent_plan_prompt}
  \end{subfigure}

  \medskip

  \begin{subfigure}[t]{\linewidth}
    \centering
    \begin{minted}[breaklines,frame=lines,framesep=2mm,baselinestretch=1.05,fontsize=\scriptsize,bgcolor=promptgray,bgcolorpadding=2mm]{text}
You are proposing one experiment for a bounded hyperparameter search.

Task: {task_description}
Objective: {objective}
Current strategy: {strategy_excerpt}
Resolved task configuration: {task_config_excerpt}
Search space: {searchable_parameters}
Current search state: {current_state}
Recent experiments: {recent_results}
Summary of earlier experiments: {earlier_summary}
Per-seed diagnostics: {seed_diagnostics}
Validation feedback, if present: {validation_feedback}

Choose exactly one listed configuration parameter and propose a bounded set of candidate values. You may select
a valid configuration from an earlier cycle as the search base.

Return one JSON object:
{
  "summary": "...",
  "search_base_selection": {
    "source": "cycle_XXX",
    "rationale": "..."
  },
  "params": [{
    "name": "path.to.parameter",
    "values": [v1, v2, v3]
  }]
}
  \end{minted}
    \caption{\emph{propose} prompt.}
    \label{fig:agent_propose_prompt}
  \end{subfigure}
  \caption{Prompt templates for the \emph{plan} and \emph{propose} stages.}
  \Description{Two framed text templates are stacked vertically. The plan template supplies the task, objective, search space, accumulated results, and per-seed diagnostics, then requests a six-part Markdown strategy. The propose template supplies the updated strategy and requests one bounded parameter search as a JSON object.}
  \label{fig:agent_prompt_templates}
\end{figure}

The curly-braced placeholders in Figure~\ref{fig:agent_prompt_templates}
denote these inputs. The \emph{plan} prompt asks the model to revise the
strategy from the accumulated evidence and choose the hyperparameter axis
for the next line search; it may also retain alternative directions for
later cycles. The \emph{propose} prompt combines the revised strategy with
the resolved task configuration and asks the model to return candidate
values for one hyperparameter as a JSON object.

Before training, the executor validates each proposal against the search space. Accepted candidates are trained and benchmarked using the configured seeds, and the results update the search state. If validation fails, the error is included in the next \emph{propose} prompt.

\subsection{Backflip Search}
\label{sec:backflip_hpo_appendix}

We use this procedure for the backflip motion-tracking task reported in the
main paper.
The initial configuration was adapted from Isaac Lab's
\texttt{Isaac-Humanoid-Direct-v0} environment~\cite{mittal2025isaac} without
additional manual tuning for the backflip task. The \texttt{\seqsplit{objective}} field was to minimize the median of the three
seed-level \textit{active training times} required to reach a success rate of
$0.9$. \textit{Active training time} measures elapsed wall-clock time
excluding initialization, checkpointing, and evaluation, all of which occur
outside the training loop. We evaluated the policy every five iterations in
32 deterministic environments. The success criterion was full-horizon
completion in at least $90\%$ of the environments. Average pose error, global
mean per-joint position error, and end-effector error were reported as
diagnostic metrics.

The 80-cycle search found its best configuration at cycle~65, reducing the
time to success from $18.77\,\mathrm{s}$ to $2.17\,\mathrm{s}$
(Figure~7 in the main paper).
Table~\ref{tab:backflip_config_change} lists its hyperparameter changes.

\begin{table}[ht]
  \centering
  \caption{Hyperparameters of the best configuration found at cycle~65 of the backflip search.}
  \label{tab:backflip_config_change}
  \small
  \begin{tabular}{@{}p{0.57\columnwidth}rr@{}}
    \toprule
    Parameter & Baseline & Selected \\
    \midrule
    Action scale & $1.0$ & $0.875$ \\
    Root target observation & False & True \\
    Motion randomization std. & $0.02$ & $0.01$ \\
    Pose reward weight & $10$ & $12$ \\
    Center-of-mass reward weight & $10$ & $8$ \\
    End-effector rotation reward weight & $2$ & $1$ \\
    Termination height & $0.8$ & $0.6$ \\
    Early-termination grace steps & $4$ & $3$ \\
    Initial policy log std. & $-2.0$ & $-1.5$ \\
    Rollout steps & $32$ & $8$ \\
    Learning epochs & $5$ & $3$ \\
    Discount factor & $0.99$ & $0.98$ \\
    GAE parameter, $\lambda$ & $0.95$ & $0.99$ \\
    Learning rate & $5.0{\times}10^{-4}$ & $6.25{\times}10^{-4}$ \\
    KL threshold & $0.008$ & $0.12$ \\
    PPO ratio clip & $0.20$ & $0.32$ \\
    Entropy coefficient & $0$ & $2.5{\times}10^{-4}$ \\
    Value-loss coefficient & $2$ & $4$ \\
    \bottomrule
  \end{tabular}
\end{table}

\paragraph{Consistency across runs.}
To assess run-to-run variation in the reported HPO result, we performed five
independent 80-cycle searches using GPT-5.5-high via Codex. The median best
active training time was $2.291\,\mathrm{s}$ with an IQR of
$0.095\,\mathrm{s}$, and all five values were between $2.031$ and
$2.449\,\mathrm{s}$.

\paragraph{Search results across LLMs.}
To evaluate the same search protocol with an open-weights LLM, we performed
five searches using GLM-5.2-high via OpenCode under the same configuration.
Across these searches, the median best time was $2.530\,\mathrm{s}$ and the
IQR was $0.571\,\mathrm{s}$, both higher than the corresponding GPT-5.5-high
results of $2.291\,\mathrm{s}$ and $0.095\,\mathrm{s}$. One GLM search,
however, found a configuration with a training time of
$1.748\,\mathrm{s}$, the fastest among all ten searches.

\begin{table}[ht]
  \centering
  \caption{Best active training times from five independent 80-cycle searches with each LLM.}
  \label{tab:agent_repeated_runs}
  \small
  \begin{tabular}{@{}lcc@{}}
    \toprule
    LLM & Median / IQR (s) & Range (s) \\
    \midrule
    GPT-5.5-high & $2.291 / 0.095$ & $2.031$--$2.449$ \\
    GLM-5.2-high & $2.530 / 0.571$ & $1.748$--$3.141$ \\
    \bottomrule
  \end{tabular}
\end{table}

\end{document}